\documentclass[epsf,aps,prb,twocolumn,nofootinbib,showpacs]{revtex4}
\usepackage{graphicx}
\usepackage{amsfonts}
\usepackage{amssymb}

\usepackage{graphicx}
\usepackage{amsfonts}
\usepackage{amssymb}
\usepackage{epsfig}
\usepackage{psfrag}
\usepackage{bm}
\newcommand{\beqn}{\begin{eqnarray}}
\newcommand{\eeqn}{\end{eqnarray}}
\newcommand{\be}{\begin{equation}}
\newcommand{\ee}{\end{equation}}

\def\beq{\begin{equation}}
\def\eeq{\end{equation}}
\def\beqn{\begin{eqnarray}}
\def\eeqn{\end{eqnarray}}

\def\s1{$s_{\alpha}$}
\def\s2{$s_{\gamma}$}
\def\s3{$s_{\delta}$}
\def\c1{$c_{\alpha}$}
\def\c2{$c_{\gamma}$}
\def\c3{$c_{\delta}$}
\def\s{Stueckelberg~}

\newcommand{\mathsym}[1]{{}}

\begin{document}
\begin{center}
\end{center}

\title{Bulk Fermi surface and momentum density in heavily doped
La$_{2-x}$Sr$_x$CuO$_4$ using high resolution Compton
scattering and positron annihilation spectroscopies}

\author{
W. Al-Sawai$^1$, B. Barbiellini$^1$, Y. Sakurai$^2$, M. Itou$^2$, P.
E. Mijnarends$^{1,3}$, R. S. Markiewicz$^1$, S. Kaprzyk$^{1,4}$, S.
Wakimoto$^5$, M. Fujita$^6$, S. Basak$^1$, H. Lin$^1$, Yung Jui Wang$^1$, 
S.W.H. Eijt$^3$, H. Schut$^3$, K. Yamada$^{6,7}$, and A. Bansil$^1$}

\affiliation{
$^1$Department of Physics, Northeastern University, Boston, MA 02115,
USA.\\
$^2$Japan Synchrotron Radiation Research Institute (JASRI), SPring-8,
1-1-1 Kouto, Sayo, Hyogo 679-5198, Japan.\\
$^3$Department of Radiation, Radionuclides $\&$ Reactors, Faculty of
Applied Sciences, Delft University of Technology, Delft, The
Netherlands.\\
 $^4$ Academy of Mining and Metallurgy AGH, 30059 Krakow, Poland.\\
 $^5$ Quantum Beam Science Directorate, Japan Atomic Energy Agency, Tokai,
Naka, Ibaraki 319-1195, Japan.\\
 $^6$ Institute of Materials Research, Tohoku University, Sendai 980-8577,
Japan.\\
$^7$ Advanced Institute of Materials Research, Tohoku University, Sendai 980-8577,
Japan.}

 \date{\today}


\begin{abstract}
We have observed the bulk Fermi surface (FS) in an overdoped
($x$=0.3) single crystal of La$_{2-x}$Sr$_x$CuO$_4$ by using Compton
scattering. A two-dimensional (2D) momentum density 
reconstruction from measured
Compton profiles yields a clear FS signature
in the third Brillouin zone along [100].
The quantitative agreement between density functional
theory (DFT) calculations and momentum density experiment suggests
that Fermi-liquid physics is restored in the overdoped regime. 
In particular the predicted FS  topology is found to be 
in good accord with the corresponding experimental data. 
We find similar quantitative agreement between the
measured 2D angular correlation of positron annihilation radiation
(2D-ACAR) spectra and the DFT based computations. However, 2D-ACAR does not give such
a clear signature of the FS in the extended momentum space in either the
theory or the experiment.

\end{abstract}
\maketitle

\section{Introduction}
High temperature cuprate superconductors \cite{DSH,BeK}
such as La$_{2-x}$Sr$_x$CuO$_4$ (LSCO)
show interesting  Mott insulator behavior
\cite{sudip,fulde} at low doping $x$ while at high doping 
these materials display Fermi liquid properties.  
\cite{DSH,proust,hussy} This fascinating phase diagram 
is complicated by the presence of various inhomogeneities such as 
stripe ordering \cite{Kivel} and possible clustering of dopants.\cite{BB} 
Many models of the overdoped metallic phase have been put forward, 
some generalizing the concept of the Fermi liquid, 
while others attempt to use Hubbard or t-J models. 
In this context, it is important to investigate the electron momentum density of 
these materials, which, when compared to predictions of theoretical models, 
can give an indication of their correctness or applicability.  

Insight into the evolution of electronic states can be obtained via Fermi
surface (FS) measurements in various doping regimes. Experimental FS work
on the cuprates has to date been limited mainly to angle-resolved
photoemission (ARPES),\cite{DSH,BeK} quantum oscillations (QO),\cite{QOsc}
and scanning tunneling spectroscopies (STS).\cite{hanaguri} However, ARPES\cite{abARPES} and STS\cite{abSTM} are surface sensitive probes. Although QO measurements probe the bulk, they provide only FS areas without giving information on the location of the FSs in momentum space. Furthermore QO require long mean free paths and large magnetic fields which could alter the ground state. For underdoped samples there is evidence of FSs distorted from
Local Density Approximation (LDA) based band structure predictions, 
involving arcs or pockets, but as doping is increased it
appears that a large, LDA-like FS becomes manifest as 
the pseudogap collapses near optimal doping. 
For an overdoped Tl-cuprate, the FS has
been found to be large and closed around $(\pi ,\pi )$,
\cite{peets07} suggesting that
the Van Hove singularity (vHS) still lies below the Fermi energy $E_F$
in good agreement with LDA calculations.\cite{bba94}

These considerations provide motivation for deploying genuinely bulk
sensitive spectroscopies for FS measurements in the cuprates. Two
such spectroscopies, which have been used extensively for this
purpose, are high-resolution Compton scattering \cite{cs1,cs2}
and two-dimensional angular correlation of positron 
annihilation radiation (2D-ACAR).\cite{acar1,acar2,acar3}
Compton scattering probes the momentum density of the many-body
electronic ground state of the system and is {\it insensitive} to
the presence of defects or surfaces in the sample.\cite{lsmo}
2D-ACAR also probes the bulk momentum density, but its
interpretation can be complicated by   positron spatial
distribution effects.\cite{manuel_acar_review,Blandin,howel} On the
other hand, the Compton scattering technique requires large single
crystals of materials having atoms with high atomic numbers
($Z\sim30$),
as is the case for all cuprates, but it suffers from the problem of a
relatively low signal from valence electrons sitting on the large
background contribution from the core electrons.

With this background, the present article reports a study of the FS and
electron momentum density (EMD) of an overdoped single crystal of
La$_{2-x}$Sr$_x$CuO$_4$ (LSCO) for the hole doping level $x$=0.3. Compton
scattering and 2D-ACAR experiments have been carried out on the same LSCO
sample, and the results are analyzed through parallel computations of the
FS, the EMD as well as the electron-positron momentum density within the
framework of DFT. The conventional picture
of the metallic state based on   Landau Fermi liquid theory is expected
to become increasingly viable with doping as the system becomes more
weakly correlated, even though the physics of the cuprates is generally
dominated by deviations from such a simple picture of electronic states.
For this reason, the overdoped LSCO provides a good starting system for
investigating the fermiology of the cuprates.

Although earlier Compton studies have provided some insight into the ground-state momentum density and electron correlation effects in LSCO,
\cite{laukkanen,science_lsco} we attempt to study the FS
of LSCO using high-resolution Compton scattering experiments.
The 2D-ACAR experiments, on the other hand, have been deployed successfully by several groups in the past for delineating the FSs of the cuprates in a number of favorable cases.\cite{manuel_acar_review}

Here, by measuring a series of high resolution Compton profiles, we
reconstruct the 2D momentum density in overdoped LSCO and identify a clear
signature of the FS in the third Brillouin zone. Moreover, the DFT-based
theoretical EMD is found to be in quantitative accord with the Compton
scattering results, indicating that the ground state wavefunction in the
overdoped system is well-described by the weakly correlated DFT picture.
We have also found a quantitative level of agreement between the measured
and computed 2D-ACAR spectra. However, the 2D-ACAR spectra do not reveal
clear FS signatures due to the well-known positron spatial distribution
effects, which can make the positron insensitive to electrons in the Cu-O
planes.

The remainder of this article is organized as follows. Section II
describes experimental details of sample preparation and of
Compton scattering and positron-annihilation experiments. Section III
provides technical details of our DFT-based EMD, FS and
electron-positron momentum density computations. Section IV
discusses momentum density anisotropy results, while Sec. V considers
the FS determination. The article
concludes with a summary of the results in Sec. VI.

\section {EXPERIMENTS}

The heavily overdoped single crystal ($x=0.30$) was grown by the traveling
solvent floating zone method. For this purpose, a powder sample was first
synthesized by the conventional solid state reaction method. It was then
shaped into feed rods under hydrostatic pressure and sintered at 1173 K
for 12 hours, and at 1423 K for an additional 10 hours. In this process,
excess CuO of 2 mol\% was added to the feed rods to compensate for the
evaporation of CuO during the high temperature process. The grown crystal
was subsequently annealed under an oxygen pressure of 3 atm at 1173 K for
100 hours. 
Superconducting quantum interference device 
(SQUID) measurements, using MPMS-XL5HG (Quantum Design, Inc.), showed 
no superconductivity down to 2 K.
Neutron diffraction studies indicated that the crystal is tetragonal
(I4/mmm) down to the lowest temperature.
The crystal was also characterized by other experiments. 
\cite{exp1,exp2,exp3}

We have measured 10 Compton profiles with scattering vectors equally
spaced between the [100] and [110] directions using the Cauchois-type
x-ray spectrometer at the BL08W beamline of SPring-8.\cite{spring8} All
measurements were carried out at room temperature. The overall momentum
resolution is estimated to be 0.13 a.u. full-width-at-half-maximum (FWHM).
The incident x-ray energy was 115 keV and the scattering angle was
$2.88$ rad. Approximately 5$\times 10^5$ counts in total were collected at
the Compton peak channel, and two independent measurements were performed
in order to check the results. Each Compton profile was corrected for
absorption, analyzer and detector efficiencies, scattering cross section,
possible double scattering contributions, and x-ray background. The
core-electron contributions were subtracted from each Compton profile. A
two-dimensional momentum density, representing a projection of the
three-dimensional momentum density onto the $a-b$ plane, was reconstructed
from each set of ten Compton profiles using the direct Fourier transform
method.\cite{fourier}\\

The 2D-ACAR was measured using the Delft University 2D-ACAR
spectrometer\cite{falub} with a conventional $^{22}$Na positron
source. The data were taken at a temperature of about $T=70$ K. To
correct for the sample shape $(5.7 \times 3.5 \times 4.5$ mm$^3)$
the data were convoluted with a gaussian of FWHM of 0.1 channel in
the $x$ direction and 1.7 channels in the $y$ direction (along which
the positrons impinge on the sample), where one
channel corresponds to 0.184 mrad. The total resolution is $1.0
\times 1.0$ mrad$^2$ FWHM (1 mrad = 0.137 a.u). The total number of
coincidences collected is $76.3 \times 10^6$ and the maximum number
of coincidences at the center is $16.7\times10^3$ counts.

The effects of superconductivity on the 2D-ACAR as well as 
Compton spectra \cite{peter} are expected to be below the 
resolution of the current  experiments. For this reason, 
we have not carried out experiments on the superconducting 
state in connection with the present study.

\section {CALCULATIONS}

Our electronic structure calculations are based on the LDA within the framework of the DFT. 
An all-electron fully charge-self-consistent semi-relativistic
Korringa-Kohn-Rostoker (KKR) methodology was used.\cite{bansil} The
crystal structure of LSCO was taken to be body-centered
tetragonal (BCT) with space group $I4/mmm$ (139) using lattice parameters $a$
and $c$ given in Ref. \onlinecite{sahrakorpi}. A non-spinpolarized
calculation
neglecting the magnetic structure was performed. Self-consistency
was obtained for $x=0$ and the effect of doping was treated
within a rigid band model by shifting the Fermi energy to
accommodate the proper number of electrons.\cite{foot2ab,footAB2,footAB3}
The results are in good agreement  
with other calculations.\cite{freeman}
The formalism for
computing momentum density, $\rho(\textbf{p})$, is discussed in
Refs. \onlinecite{mijnarends1,mijnarends3,mijnarends4,mijnarends5}.
The EMD is calculated according to the formula \be
\begin{array}{c}
\rho(\textbf{p})=\sum_{i}
n_i(\textbf{k})\mid\int
\exp(-i \textbf{p}.\textbf{r})\psi_i(\textbf{r})d\textbf{r}\mid^2\\
\label{eq1}
\end{array}
\ee and the ACAR spectrum is computed as \be
\begin{array}{c}
\rho^{2\gamma}(\textbf{p})=\sum_i n_{i}(\textbf{k})\mid\int
\exp(-i
\textbf{p}.\textbf{r})\psi_i(\textbf{r})\phi_+(\textbf{r})d\textbf{r}\mid^2,\\
\label{eq2}
\end{array}
\ee where $\psi_i$ denotes the electronic wave function, $\phi_+$
the positron wavefunction, and $\textbf{p}=\textbf{k}+\textbf{G}$ with $\textbf{G}$ 
a reciprocal lattice
vector. $n_{i}(\textbf{k})$ is the occupation function \cite{agp}
which in the independent particle model equals
$1$ if the electron state $i$ is occupied and $0$ when it is empty, and the 
summation extends over the occupied $\textbf{k}$ states. In the
$\rho^{2\gamma}(\textbf{p})$ calculation, we have neglected the
enhancement factor for the annihilation rate.\cite{gga} The
inclusion of enhancement effects is crucial for the calculation of
lifetimes but is well-known \cite{Blandin,icpa_bba92}
to be not very important for discussing
questions of bonding and FS signals in momentum density. 
The momentum densities are calculated on a momentum mesh with step
($\delta p_x,\delta p_y,\delta p_z)=2\pi (1/32a,1/32a,1/4c)$. 
The momentum is expressed in atomic units defined by $1/a_0$
where $a_0$ is the Bohr radius.
The calculations include contributions from both the filled
valence bands and the conduction-band which gives rise to the FS in
LSCO. To study the electronic structure of the system, we consider
two quantities of interest, 2D-EMD and 1D-EMD, which are the
projection of EMD in 2D and 1D, respectively, given by:
\begin{equation}
\rho^{2d}(p_x,p_y)= \int \rho(\textbf{p})dp_z 
\label{eq3}
\end{equation}
 and 
\begin{equation}
\rho^{1d}(p_x)=\int \int \rho(\textbf{p})dp_y dp_z.\\
 \label{eq4}
\end{equation}

\begin{figure}
 \includegraphics[width=8.0cm]{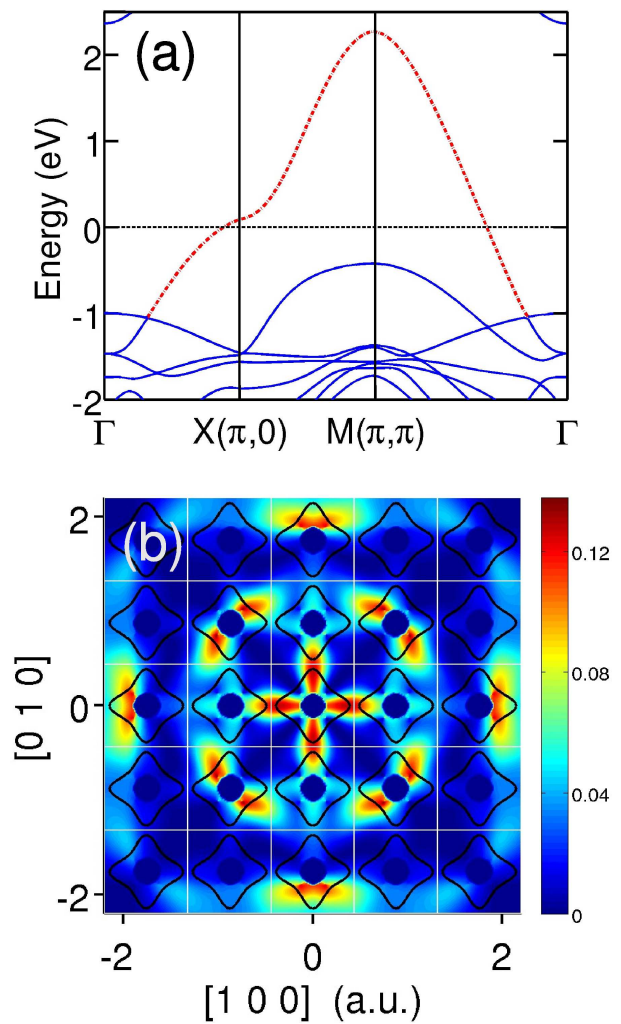}
 \caption{(Color online)
 (a) Band structure of LSCO near the Fermi level.
 The CuO$_2$ band is shown as a red dotted line.
 (b) Calculated $\rho^{2d}_{x^{2}-y^{2}}(p_x,p_y)$  
 is shown together with the FS sections at $(k_{z}=0)$ for the doping level 
 $x=0.3$. The color scale is in units of $\rho^{2d}$(0,0).
 The grid represents Brillouin zones with a size of $2 \pi/a$, where $a$
 is the lattice constant.}
 \label{fig0}
 \end{figure}

 The calculated band structure of LSCO ($x=0.3$) 
 near the Fermi level is illustrated 
 in Fig.~\ref{fig0}(a). 
 The band closest to the Fermi level is shown by the red dotted curve.
 This CuO$_2$ energy band is dominated by copper-oxygen $d_{x^{2}-y^{2}}-p_{x,y}$ orbitals.
 The present calculation for $x$=0.3 predicts 
 that the vHS is above the Fermi energy, so the FS has a diamond shape
 closed around the $\Gamma$ point.
 Figure~\ref{fig0}(b) shows the 2D momentum density contribution of
 the $x^{2}-y^{2}$ band, $\rho^{2d}_{x^{2}-y^{2}}(p_x,p_y)$,  
 together with the FS sections at $(k_{z}=0)$ for the doping level 
 $x=0.3$ mapped periodically throughout the momentum space.
 The Brillouin zones are visualized in the figure by a grid of $2 \pi/a$, where $a$
 is the lattice constant. The momentum density acts as a matrix element on the FS map.  
 Thus, since
 $\rho^{2d}_{x^{2}-y^{2}}(p_x,p_y)$ 
 has strong amplitudes in the third Brillouin zones, the FS could
 be more easily detected there.

The positron annihilation and Compton scattering experiments probe all
the electrons in the system. However, core and
semi-core electrons
give isotropic distributions while the anisotropy
of the spectra is produced by electrons
near the Fermi energy.
Therefore, we can concentrate on analyzing the residual 
anisotropy after
subtraction of the isotropic part. We consider an anisotropy with
$C_{4v}$ symmetry, given by\cite{c4v}:
\begin{equation}
\begin{array}{c}
A_{C_{4v}}^{2d}(p_x,p_y)=
\rho^{2d}(p_x,p_y)-
\rho^{2d}(\frac{p_x+p_y}{\sqrt{2}},
\frac{p_x-p_y}{\sqrt{2}}).\\
\label{eq5}
\end{array}
\end{equation}
In order to enhance FS signals, it is also possible 
to calculate another anisotropy obtained by subtraction of a 
smooth cylindrical average of the distribution $\rho^{2d}(p_x,p_y)$ 
defined by
\begin{equation}
A^{2d}(p_x,p_y)=\rho^{2d}(p_x,p_y)-S(\sqrt{p_x^2+p_y^2}),
\label{eq7}
\end{equation}
where $S$ is a smoothed cylindrical average of $\rho$, 
in which the original spectrum is averaged
over rotation angles from 0$^\circ$ to 45$^\circ$ in steps of 1$^\circ$.

\begin{figure}[t]
\includegraphics[width=9.2cm,height=5.0cm]{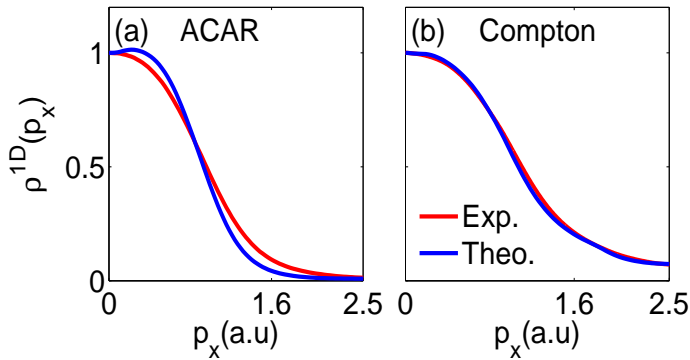}
\caption{(color online) Theoretical and experimental profiles
$\rho^{1d}(p_x)$ for (a) ACAR and (b) Compton scattering along
the [100] direction. All spectra are normalized 
to $\rho^{1d}$(0). } \label{fig1}
\end{figure}

\section {Momentum Density Anisotropy}
\begin{figure}[t]
\includegraphics[width=9.2cm,height=5.0cm]{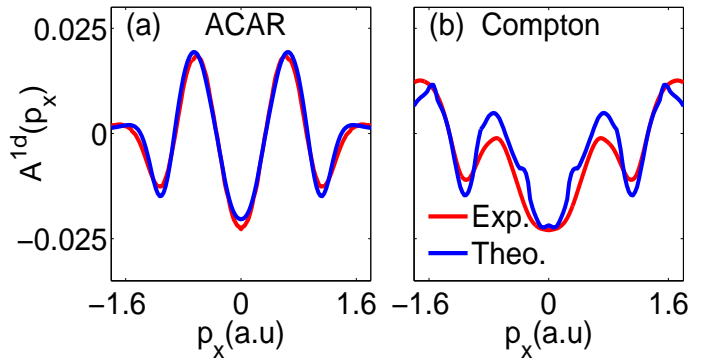}
\caption{(color online) Theoretical and experimental $A^{1d}(p_x)$
for (a) ACAR and (b) Compton scattering} \label{fig4}
\end{figure}

We start by comparing the experimental 1D profiles based on Compton scattering
and ACAR measurements with the corresponding theoretical predictions.
Figure~\ref{fig1} shows a good level of accord between theory and
experiment, especially for the Compton results. 
Further insight is gained by considering the 1D-projection of the
anisotropy of Eq.~\ref{eq5} given by 
 \begin{equation}
A^{1d}(p_x)=\int_{p_{a}}^{p_{b}} A^{2d}_{C_{4v}}(p_x,p_y) dp_y~,
 \label{eq6}
 \end{equation}
where $\Delta p=p_{b}-p_{a}$ is the momentum range over which the
projection is taken.\cite{note2}
The profile $A^{1d}(p_x)$, shown in Fig.~\ref{fig4}, 
is also equal to the difference of profiles \cite{lsmo} 
between two crystallographic directions 
[100] - [110] 
and it can be compared to a similar Compton profile anisotropy 
measured by Shukla {\em et al.} at a lower hole doping.\cite{shukla99}
The amplitude of the theory is the same as that in the experiment, 
while in Ref. \onlinecite{shukla99}
the theoretical	$A^{1d}(p_x)$ had to be	
scaled down by a factor of $1.4$ 
to obtain agreement with experiment.
We note that theory predicts that
the main contribution of correlation effects 
is an isotropic redistribution of the momentum density,\cite{lp} so
that the amplitudes of oscillation in $A^{1d}(p_x)$ become significantly reduced 
in the strongly correlated system.  Hence our quantitative agreement between
theory and experiment suggests that the $x=0.30$ doped regime
is consistent with Fermi liquid physics and that correlation effects 
modifying the anisotropy \cite{shukla99,bba_platzman} are relatively weak. 

\begin{figure}[t]
\includegraphics[width=8.2cm]{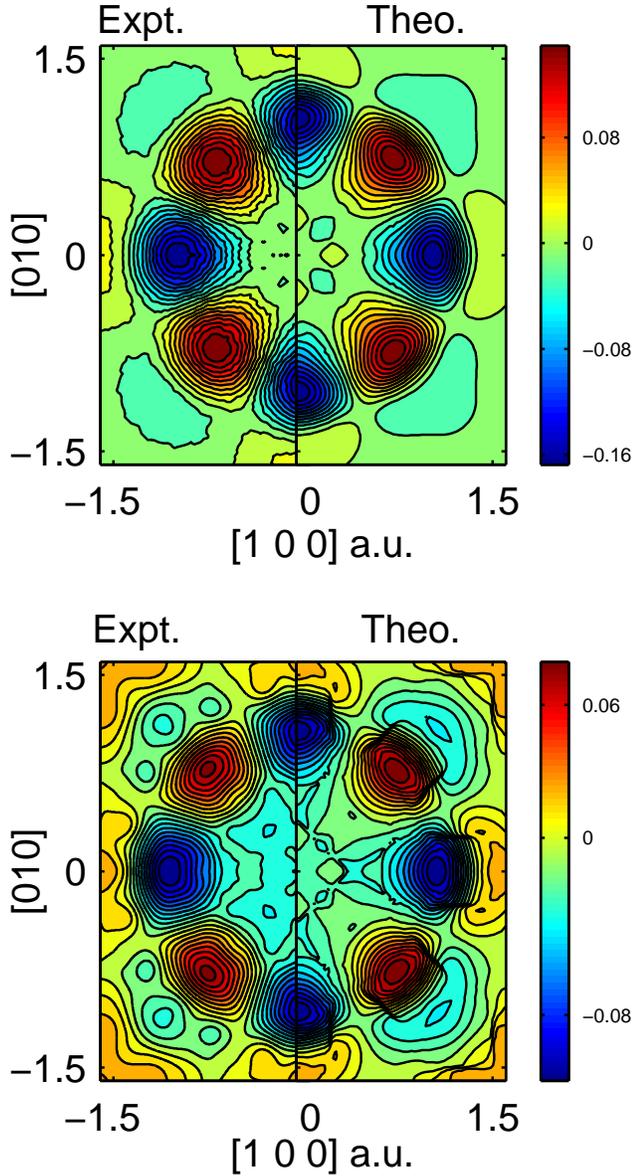}
\caption{(color online) Top: (left) Experimental and 
(right) theoretical ACAR
$C_{4v}$ anisotropy.
Bottom:(left) Experimental and (right) theoretical 
$C_{4v}$ anisotropy (Compton).
The color scale is in units of $\rho^{2d}$(0,0).} 
\label{fig23}
\end{figure}

In order to focus on
the Cu-O band contribution, Fig.~\ref{fig23} presents the 2D
$C_{4v}$ anisotropy distributions in the ($p_x,p_y$) plane for positron
annihilation and Compton scattering spectra, respectively.~\cite{note1} 
The anisotropy of the ACAR \cite{note1} in Fig.~\ref{fig23}
can be modeled by a molecular orbital method \cite{turchi} involving the
overlap of the positron wavefunction with Cu 3$d$ states hybridized with 
O 2$p$ states. 
For an atomic orbital, the momentum density has the same point symmetry as 
the corresponding charge density. 
This result carries over to molecular states 
\cite{turchi} and is equally applicable 
to solid-state wave functions.\cite{harthoorn78}
The Compton scattering anisotropy maps are
very similar overall to the positron-annihilation results,
except that the Compton spectra extend to significantly higher momenta.
This is expected since in the ACAR case the tendency of the positron to
avoid positively charged ionic cores has the effect of suppressing higher
momentum components of the EMD produced by the core and the localized 
valence electrons.
Figure~\ref{fig23} clearly shows that the theory reproduces
most of the structure observed in the measured Compton as well as
positron-annihilation distributions, including the momenta at which
various features are located. The fourfold symmetry of the spectra is a
consequence of the symmetry of the underlying body-centered tetragonal
lattice. 

Therefore, an important finding 
of our work is that the correlations that ordinarily
reduce the Compton anisotropy 
amplitudes are no longer effective at this
doping.  Similar Fermi liquid behavior has been reported 
in studies of overdoped Tl-cuprates.\cite{DSH,proust,hussy}
Moreover, the trend of the weakening of correlation effects  
with doping is also consistent with the 
changes observed in x-ray absorption \cite{towfiq,peets09} and 
photoelectron spectroscopies.\cite{schneider}

\section {FERMI SURFACE RESULTS}
\begin{figure}[t]
\includegraphics[width=8.0cm,height=4.5cm]{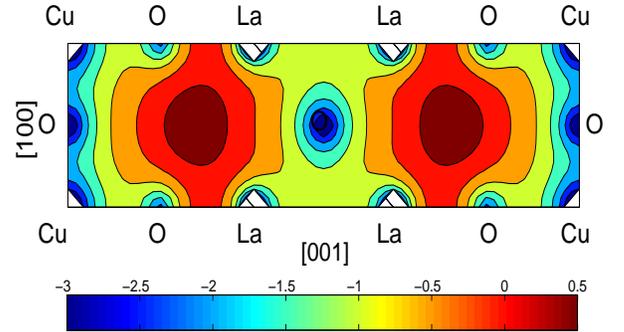}
\caption{(color online) Positron density for LSCO in the
(010) plane. 
The positron wave-function is normalized in the LSCO
unit cell and a logarithmic density scale 
(with integer numbers as units) is used in order to enhance
regions with low positron density.
The atomic positions (La, Cu, O) set the scale for 
the $x$ and $y$ axes.} \label{fig5}
\end{figure}

We comment briefly on the positron results first. Although the
electron-positron momentum density measured in a positron-annihilation
experiment contains FS signatures, the amplitude of such signatures is
controlled by the extent to which the positron wavefunction overlaps with
the states at the Fermi energy. 
Figure~\ref{fig5} shows the positron density distribution in a planar section
of the LSCO unit cell. As in the calculation by Blandin {\em et al.} 
\cite{Blandin}, our result indicates that the positron does 
not probe well the FS contribution of the Cu-O planes in
LSCO. Moreover, positron-annihilation favors FS contributions involving O 2$p$ 
rather than Cu 3$d$ states because the positron wavefunction overlaps more with 
extended O $2p$  states compared to the more tightly bound Cu 3$d$ states. 

Indeed, we see little evidence of FS signatures
in either the computed or the measured ACAR distributions 
Fig.\ref{fig4}~(a).
A more favorable case  for the 2D-ACAR distribution
is provided by the YBa$_2$Cu$_3$O$_7$ cuprate superconductor,
where the 1-dimensional ridge FS has a two-fold symmetry which distinguishes
it from important four-fold symmetry wave function 
effects.\cite{haghighi,hoffmann}  On the other hand, the Compton scattering
calculation reveals a clear Fermi surface feature near a momentum $(1.5,1.5)$ a.u., and equivalent positions,
Fig.~4(b).  Strikingly, a similar feature is seen in the experimental spectrum on the left-side of the figure.
To see this more clearly, we apply a Lock-Crisp-West folding procedure, Fig. 6.

The Lock-Crisp-West (LCW) theorem\cite{LCW} can be used 
both in Compton scattering 2D-EMD and in the case of 2D-ACAR data
to study the non isotropic features of the momentum
density by folding the data into a single central Brillouin zone. This technique can enhance FS discontinuities
(`breaks') by coherently superposing the umklapp terms.\cite{Blandin} However, the LCW folding can also
artificially enhance errors in the experimental data.\cite{bba94} Thus, to more clearly expose FS features 
in the data, we consider the cylindrical anisotropy defined by Eq.~\ref{eq7}.
\begin{figure}[t]
\includegraphics{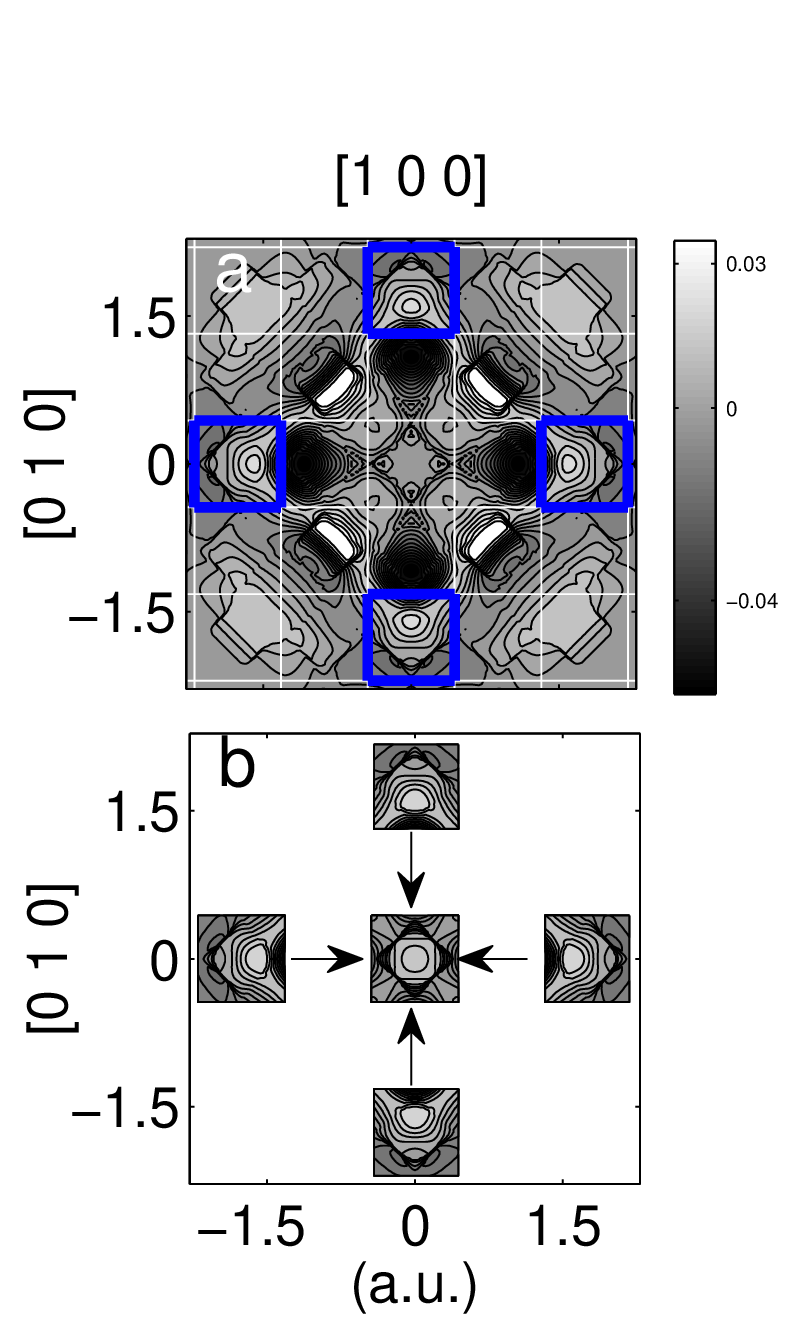}
\caption{(color online)  (a) Cylindrical anisotropy for theoretical Compton scattering momentum density.
The white lines define the Brillouin zones, while blue squares indicate a family of zones where the FS 
features are particularly strong. 
(b) The blue squared regions are isolated from the rest of the spectra and folded back to the central region.
The color scale is in units of $\rho^{2d}$(0,0).}
\label{fig6}
\end{figure}
Since the subtracted function is smooth and slowly varying this procedure does not
contribute to, nor create, new structures in the spectrum remaining after subtraction.

\begin{figure}[t]
\includegraphics{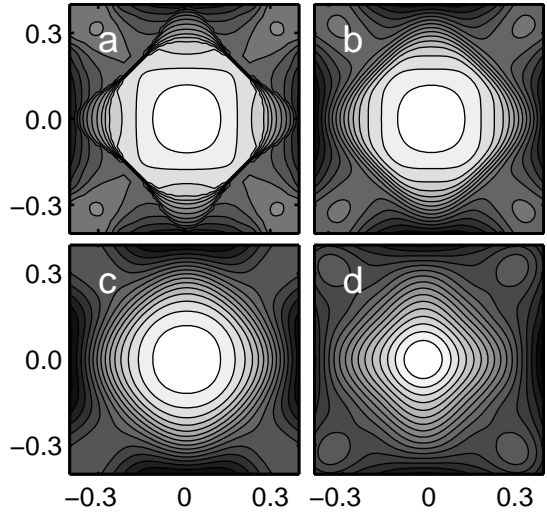}
\caption{Result of the 'partial folding' procedure: 
(a) Non convoluted Theory; 
(b) Theory convoluted with experimental resolution $0.07$ a.u..
(c) Theory convoluted with experimental resolution $0.13$ a.u.;
(d) Experiment. The white color corresponds to a weight of unity (occupied) 
while the black color corresponds to zero (unoccupied).}
\label{fig7}
\end{figure}

Figure \ref{fig6}~(a) shows the cylindrical anisotropy of the theoretical 2D-EMD spectrum.
The momentum density is represented in the extended zone scheme. Because the FS is periodic, 
a complete FS must exist in each Brillouin zone,
but with its intensity modulated by matrix element  effects, as in Fig.~1(b). 
For a predominantly $d$-wave FS, 
the matrix element effects will strongly suppress spectral
weight near $\Gamma$, so the FS breaks are most clearly seen in higher Brillouin zones. These FS breaks appear superimposed on the momentum
density in the form of discontinuities which can occur in any Brillouin zone. In Fig.\ref{fig6}~(a), the Fermi surface breaks are the regions where
the contours run closely together so that the electron momentum density varies rapidly at these locations.
Figure~\ref{fig6}~(a) shows the calculated Fermi breaks in several Brillouin zones. In particular, in the third  
zones framed with blue squares, the arc-like features are theoretically predicted FSs associated with Cu-O planes. 
Due to the tetragonal symmetry, a rotation
of the spectrum by $\pi/2$ will generate symmetry-related regions (blue squares) with equivalent strong FS features.
These regions are isolated in Fig.\ref{fig6}~(b).  By performing a `partial folding', that is, folding only these
regions back into the first Brillouin zone, we produce a full FS, 
where strong matrix element effects are substantially circumvented.  
The resulting
FS map is shown in the center of Fig.\ref{fig6}~(b), and again on a larger scale in Fig.\ref{fig7}~(a). Applying the same procedure
to other Brillouin zone regions produces similar results.  For instance, the four regions along the diagonal neighboring
the central region can also yield full FS information, but here it is superimposed on strong momentum density features.

In Figure~\ref{fig7} we compare 
the theoretical FS obtained via the aforementioned `partial folding' procedure 
to the corresponding experimental result.
Figure~\ref{fig7}~(a) 
shows that the partial backfolding technique produces
the FS of correct size and topology.  
The same procedure when applied to the experimental spectrum
yields results of Fig.\ref{fig7}~(d).
To compare with theory, we have convoluted the spectrum 
in Fig.\ref{fig7}(a) 
with $0.07$ a.u and  $0.13$ a.u. resolutions, 
leading to the distributions of Figs.\ref{fig7}~(b)~and~(c) respectively.
The latter is close to the actual experimental resolution used in this work. 

\begin{figure} [t]
\includegraphics[width=7.2cm,height=7.0cm]{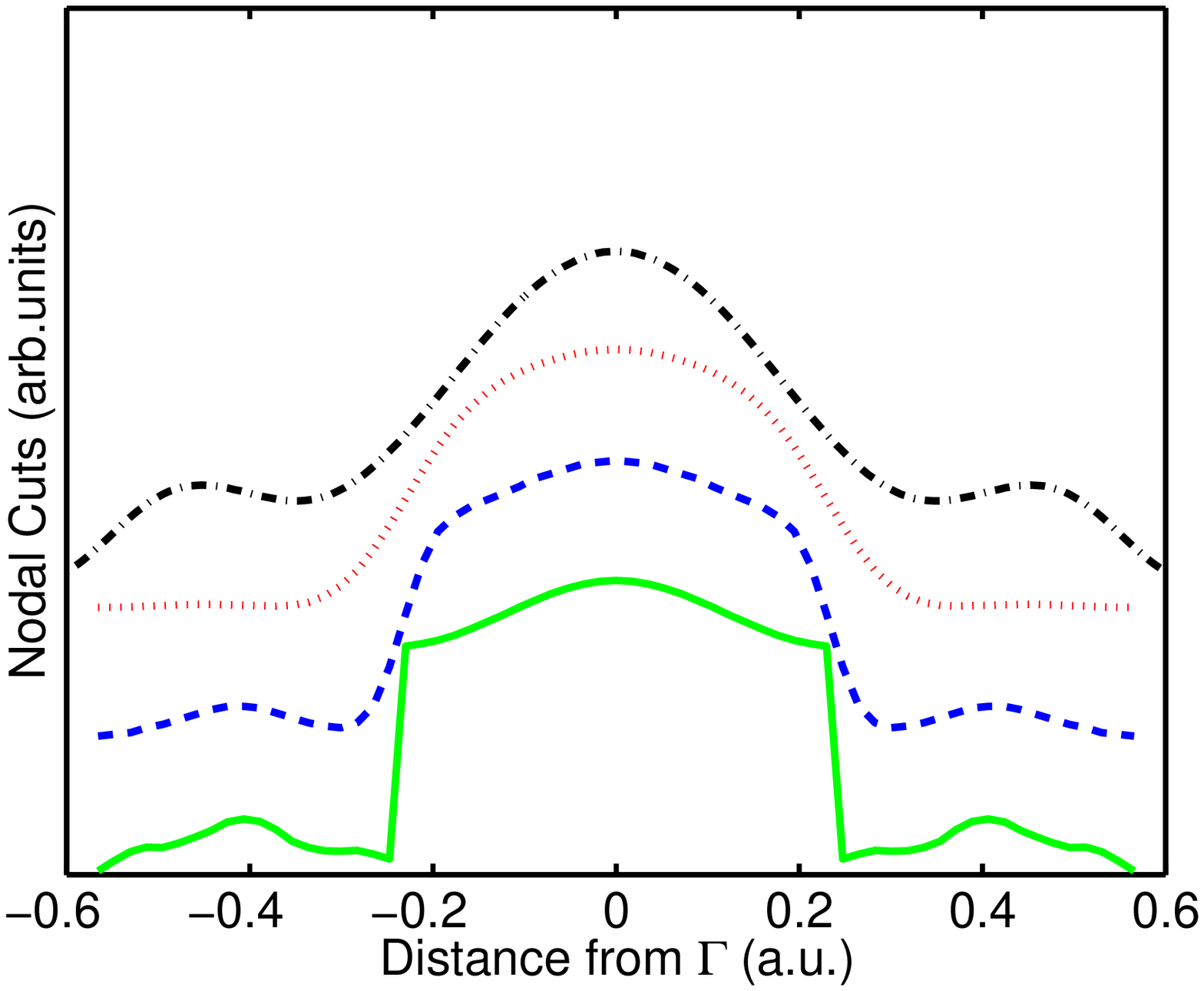}
\caption{(color online)
Cuts of the distributions
in Fig.~\ref{fig7}
along the nodal direction.
The amplitudes of theory and experiment are compared
on the same scale. 
Curves are offset with respect to one another
for clarity.
From the bottom to the top, 
the first curve is the non-convoluted theory, 
the second is the convoluted theory with a resolution $0.07$ a.u.,
the third one is the convoluted theory with a resolution $0.13$ a.u.
and the fourth curve is the experimental data.}
\label{fig8m}
\end{figure}

By comparing the four frames in Fig.~\ref{fig7}, one can see that the
experimental spectrum in Fig.~\ref{fig7}~(d) shows 
FS structures consistent with the LDA theory shown 
in Fig.~\ref{fig7}~(a).  
Clearly the FS discontinuity along the nodal direction is
smeared when including resolution broadening
as shown in Fig.\ref{fig7}~(b) and Fig\ref{fig7}~(c ).
The best overall agreement between theory and experiment 
lies between Fig.~\ref{fig7}~(b) and Fig\ref{fig7}~(c ).
This is also confirmed by the cuts taken 
along the nodal direction plotted in Fig.\ref{fig8m}. 
The remaining discrepancy between experiment and theory could 
result from intrinsic or extrinsic
inhomogeneity effects such as the appearance of local ferromagnetic
clusters about concentrated regions of dopant atoms,\cite{BB}
which have been neglected in the present simulations. 
Our conclusion is that the closed FS for $x=0.30$ predicted by our LDA calculation is consistent 
both with the present Compton scattering data and with surface-sensitive ARPES
results.\cite{Yoshida}

\section{CONCLUSIONS}
We have performed momentum density measurements 
on a high quality overdoped LSCO sample using both Compton scattering and 2D-ACAR. 
First principles calculations were also performed for the corresponding spectra.
The quantitative agreement between the calculations and the experiment for ACAR as well as
EMD anisotropies suggests that $x=0.3$ 
overdoped LSCO can be explained within the conventional Fermi-liquid theory.
Nevertheless, a FS signal was only clearly observed by Compton scattering in
the third Brillouin zone along [100]. Our FS analysis
confirms previous ARPES FS measurements \cite{Yoshida} showing an electron-like FS 
in the overdoped regime. This validation is important since  
we provide via deep inelastic x-ray scattering experiments a truly 
bulk-sensitive image of momentum density maps of electrons near the Femi level. 
In general, this momentum density information is difficult to extract 
from ARPES experiments due to difficulties associated with matrix element 
effects and the well-known surface sensitivity of ARPES.\\ 

{\bf Acknowledgements} This work is supported by the USDOE
grants No. DE-FG02-07ER46352 and No. DE-FG02-08ER46540 (CMSN) 
and benefited from the allocation of supercomputer time at NERSC 
and Northeastern University's Advanced Scientific Computation Center (ASCC),
and the Stichting Nationale Computer faciliteiten (National Computing
Facilities Foundation, NCF). 
The work at JASRI was supported by a Grant-in-Aid for Scientific Research
(nos. 18340111 and 22540382) from the Ministry of Education, Culture, Sports, Science, and Technology (MEXT), Japan, and that
at Tohoku University was supported by a Grant-in-Aid for Scientific Research (nos. 16104005, 19340090 and 22244039) from the 
MEXT, Japan.
The Compton scattering experiments were
performed with the approval of JASRI (Proposals, 2003B0762,
2004A0152, 2007B1413, 2008A1191).


\begin{thebibliography}{999}
\bibitem{foot2ab}The rigid band picture\cite{footAB2} is expected to be reasonable for doping away from the cuprate planes. It will be interesting to examine doping effects via first principles approaches \cite{footAB3}.
\bibitem{note2}
The limits of integration are within the momentum interval [-2.5, 2.5] a.u.
\bibitem{note1}
Our experimental ACAR spectra are in reasonable accord with 
earlier studies Refs. \onlinecite{Blandin,howel}.
\end{thebibliography}
\end{document}